\documentclass[prc,twocolumn,preprintnumbers,superscriptaddress,showpacs,nofootinbib]{revtex4-1}
\usepackage{graphicx,times}
\usepackage{amsmath,amsfonts,amssymb}
\usepackage{color}
\usepackage{psfrag}

\begin{document}

\title{Theoretical study of the $KK\bar K$ system  and dynamical generation of the $K(1460)$ resonance}
\author{A. Mart\'inez Torres}
\affiliation{
Yukawa Institute for Theoretical Physics, Kyoto University,
Kyoto 606-8502, Japan}
\author{D. Jido}
\affiliation{
Yukawa Institute for Theoretical Physics, Kyoto University,
Kyoto 606-8502, Japan}

\author{Y. Kanada-En'yo}
\affiliation{Department of Physics, Kyoto University, Kyoto 606-8502, Japan}

\preprint{YITP-11-19}

\date{\today}

\begin{abstract}
The $K K\bar K$ system is investigated with a coupled channel approach based on solving the Faddeev equations considering the $KK\bar K$, $K\pi\pi$ and $K\pi\eta$ channels and using as input two-body $t$-matrices that generate dynamically the $f_0(980)$ and $a_0(980)$ resonances. In the present calculation, a quasibound state around $1420$ MeV with total isospin $I=1/2$ and spin-parity $J^\pi=0^-$ is found below the three kaon threshold. This state can be identified with the $K(1460)$ resonance listed by the Particle Data Group. 
We also study the $KK\bar K$ system in a single channel three-body potential model with two-body effective $KK$ and $K\bar K$ interactions, in which the $K\bar K$ interaction is adjusted to reproduce the properties of the $f_0(980)$ and $a_0(980)$ resonances as $K\bar K$ bound states, obtaining a very similar result to the one found in the Faddeev approach. 
\end{abstract}

\maketitle

\section{Introduction}
The study of systems made by mesons and baryons and the interpretation of the states found in them is one of the challenging issues in theoretical
as well as in experimental hadron and nuclear physics. Some historical examples, which even now continue to be  of current interest, are: the study of the $\bar K N$ interaction and the formation  of the $\Lambda(1405)$ as a quasibound state \cite{da}, the searching of bound states in the $\bar K NN$ system \cite{nogami}, the possibility that some scalar resonances, like the $f_0(980)$ and $a_0(980)$, could be considered as hadronic molecular states of a system made by a $K$ and a $\bar K$ \cite{wein} and the consideration of the scalar nonet as $q\bar q$ states with a meson-meson admixture \cite{jaffe,eef}.

In the last decade, the study of few hadron systems within effective field theories has represented a step forward in the understanding of the properties of many
hadron resonances and bound states. In particular, the use of effective chiral Lagrangians to describe the meson-baryon and meson-meson
interactions implemented with unitarity in coupled channels has shed new light on the nature of several meson and baryon states like the ones studied in Refs.~\cite{da,wein} (see Refs.~\cite{oller2,kaiser}). For example,   
in the meson sector for strangeness $S=0$, within unitarity chiral theories,
the $f_0(980)$ and $\sigma(600)$ states can be understood as dynamical generated states in the $K\bar K$ and $\pi\pi$ interactions in s-wave, while the $a_0(980)$ resonance gets dynamically generated in the $K\bar K$ and $\pi\eta$ channels \cite{oller2,oller5}. 
The use of unitarity chiral theories has been also successful in the description of the meson-baryon interaction. For instance, for strangeness $S=-1$, the study of the $\bar{K}N$ and coupled channels system in $s$-wave has revealed the dynamical generation of the $\Lambda(1405)$ resonance with a double pole structure: one pole, around 1426 MeV and with a width of 16 MeV, which couples strongly to the $\bar KN$ channel, and another one, with a mass around 1390 MeV and a width of 66 MeV, which couples more to the $\pi\Sigma$ channel \cite{oller4,jido3,jido2}. It is important to 
emphasize that the $\Lambda(1405)$ can be described as 
a dynamically generated resonance without introducing an explicit pole term~\cite{Hyodo:2008xr}. 
In case of $S=0$, the investigation of the $\pi N$ system  and coupled channels in s-wave revealed significant contributions of the $K\Sigma$ and $K\Lambda$ component to the $N^*(1535)$ resonance~\cite{Kaiser:1995cy,inoue,nieves}. 

In all these systems the kaons play an important role in the dynamics due to their heavy mass (as compared to the pion) and Nambu-Goldstone boson nature. As mentioned above, the $\bar K N$ and $K\bar K$ interactions are strongly attractive in s-wave, and the fact that the kaon has a mass around 3.5 times bigger than the pion makes the s-wave interactions involving kaons more effective than those involving  pions, especially around the threshold energy. In addition, knowing that the typical kaon kinetic energy in the bound systems estimated by the hadronic interaction range is small in comparison with the kaon mass, one may treat the kaons in few-body systems within non-relativistic potential models.  

Recently, special interest has developed for few-body systems constituted by one or more kaons. For example, the $\bar{K}NN$ system has been the object of thorough studies ~\cite{yama,nina,ikeda,dote,Wycech:2008wf} and all of them indicate the presence of a quasibound state 
with a large width. Baryonic systems with two kaons, like $K\bar K N$ and $\bar K\bar K N$, were also investigated 
in Refs.~\cite{jido, enyo} with a single channel variational method and a new $N^*$ state around 1910 MeV with $J^\pi=1/2^+$ was predicted in the first case, while a very weakly bound state was found in the second case.  
A peculiar feature of these states is that they can be considered as loosely bound systems in which the identity of the constituent hadrons is kept and, thus, can be regarded as hadronic molecular states. Reflecting this fact, the inter-hadron distances are compatible with typical nucleon distances in nuclei and the size of the $K\bar K N$ system is as large as $^{4}$He \cite{jido}.
The $\Lambda(1405)$ resonance described in the unitary chiral model  also has a  larger spatial size than a typical  baryon in its ground state~\cite{Sekihara:2008qk}.

The $K \bar K N$ system was also investigated solving the Faddeev equations to obtain the three-body $T$-matrix for the system and also resulted in the finding of a $N^*$ resonance around 1920 MeV \cite{mko3,mj1}, confirming in this way the prediction done in Ref.~\cite{jido}. The approach employed in Refs.~\cite{mko3,mj1} is based on the idea of extending the unitary chiral models successfully used to study two hadron interactions to the investigation of three-hadron systems using the Faddeev equations. This formalism has been applied to study many different three-body systems made of mesons and baryons and has brought out, for the first time, the three-body nature of several resonances \cite{mko1,mko2,mko3,mko4,mko5,mko6,mko7,mj1}. For example, the study of systems like $\pi \bar K N$, $\pi\pi \Sigma$, $\pi\pi N$, $\phi K\bar K$, etc., has revealed the dynamical generation of  all the known $1/2^+$ baryon states listed by the Particle Data Group (PDG) \cite{pdg},  some $N^*$ resonances, like the $N^*(1710)$ \cite{mko2, mko3}, as well as some of the recent meson states observed in the experiments, like $\phi(2170)$ \cite{mko4}, $Y(4260)$ \cite{mko5}.

Going back to the kaonic systems, one can summarize that so far the well studied few-hadron systems constituted by anti-kaons and nucleons are: $\bar K N$, $\bar K NN$, $\bar K\bar K N$ and $K\bar K N$. Of the different states generated in these systems, the most peculiar probably are the $\Lambda(1405)$, due to its double pole nature, and the $N^*(1910)$, which is formed when the $\Lambda(1405)$ is dynamically generated in the $\bar{K} N$ subsystem and coupled channels and, at the same time,  the $K\bar K$ subsystem and coupled channels give rise to the $f_0(980)$ or $a_0(980)$ states. As shown in Refs. \cite{jido, mko3,mj1}, the attraction present in the $\bar K N$ subsystem in isospin zero and $K\bar K$ in isospin zero or one is strong enough to compensate the repulsion in the $KN$ subsystem and, thus, form a $K\bar K N$ bound state. 

The question which arises now is what will happen if instead of a nucleon we add a kaon to the $K\bar K$ system. Now, the different $K\bar K$ interactions can lead to the simultaneous presence of the $a_0(980)$ and $f_0(980)$ resonances in two of the subsystems, such that these interactions could be strong enough to overcome the $KK$ repulsion and form a $KK\bar K$ bound state or resonance. Recently, two-body systems of $f_0(980) K$ and $a_0(980) K$  have been studied 
\cite{ollerroca} using an extended version of the formalism developed in  Ref.~\cite{Luis} and a resonant peak is found at 1460 MeV with 100 MeV of width and total isospin $1/2$. However, some remarks concerning the model used in Ref.~\cite{Luis} have been made in Ref.~\cite{Javi}.

In this work, we study the possibility to form a three-body state in the $K K\bar K$ system by means of the three-body Faddeev formulation developed in Refs.~\cite{mko1,mko2,mko3,mko4,mko5,mko6,mko7,mj1}. 
One of the advantages of the present three-body formulation is that 
all the parameters involved in the approach are related to the two-body subsystems (typically a cut-off or a subtraction constant 
to regularize the two-body loops). Thus, 
we can focus on the investigation of the three-body hadron dynamics 
without introducing new adjustable parameters. 

The paper proceeds as follows: In Sec.~\ref{For} we introduce briefly the formalism employed
to determine the two-body $t$-matrices of the different subsystems and the method used to calculate the three-body
$T$-matrix of the $KK\bar K$ system and coupled channels. In Sec.~\ref{Re} we present the results obtained from the Faddeev
approach and we compare them with the ones of the nonrelativistic single channel $KK\bar K$ potential model calculation (Sec.~\ref{sec:pot}), which
is based on the work of Refs.~\cite{jido,enyo}, and which provides a simple physical picture of the $KK\bar K$ quasibound state found. Finally, in
Sec.~\ref{Co} we draw some conclusions.

\section{Formalism}\label{For}

To study the $KK\bar K$ system, we first need to determine the two-body $t$-matrices which describe the $KK$ and $K\bar K$ interactions. These
two-body amplitudes are calculated by solving the Bethe-Salpeter equation in a coupled channel approach and using the on-shell
factorization method~\cite{oller2}, in which the $t$-matrix for the system read as
\begin{equation}
t=(1-V\tilde{G})^{-1}V, \label{BS}
\end{equation}
where the interaction kernel $V$ corresponds to the lowest order amplitude obtained from chiral Lagrangians ~\cite{oller2}.  In Eq.~(\ref{BS}), $\tilde{G}$
represents the loop function of two pseudoscalar mesons and we calculate it using the dimensional regularization scheme of Ref.~\cite{oller5}
\begin{align}
\tilde{G}_r=&\frac{1}{16\pi^2}\Bigg\{a_r(\mu)+\ln\frac{m^2_{1r}}{\mu^2}+\frac{m^2_{2r}-m^2_{1r}+E^2}{2E^2}\ln\frac{m^2_{2r}}{m^2_{1r}}\nonumber\\
&+\frac{q_r}{E}\Bigg[\ln\Big(E^2-(m^2_{1r}-m^2_{2r})+2q_rE\Big)\nonumber\\
&+\ln\Big(E^2+(m^2_{1r}-m^2_{2r})+2q_rE\Big)\nonumber\\
&-\ln\Big(-E^2+(m^2_{1r}-m^2_{2r})+2q_rE\Big)\nonumber\\
&-\ln\Big(-E^2-(m^2_{1r}-m^2_{2r})+2q_rE\Big)\Bigg]\Bigg\},\label{g}
\end{align}
with $\mu$ a regularization scale and $a_r(\mu)$ a subtraction constant for the channel $r$. Following  Ref.~\cite{oller5}, we consider $\mu =1224$ MeV  and a
value for $a_r(\mu)\sim-1$. These parameters are fixed  to reproduce the observed two-body phase shifts and inelasticities for the $K\bar K$ system and coupled channels as done in Ref.~\cite{oller2,oller5}. For the $KK$ system we have assumed the same values for $\mu$ and $a_r$. In Eq.~(\ref{g}), $E$ is the total energy of the two-body system, $m_{1r}$, $m_{2r}$ and $q_r$ correspond, respectively, to the masses and the center of mass momentum of the two pseudoscalars present in the channel.  

The solution of Eq.~(\ref{BS}) is obtained by taking into account all possible two-body channels of two mesons ($\pi$, $\eta$, $K$, $\bar K$)  which couple to $ K\bar K$, $K\pi $, $K\eta$, and $KK$, except for the $\eta\eta$ channel, whose effect in the $K \bar K$ amplitude is negligible~\cite{oller2}. 
The $K\bar{K}$ and $\pi\pi$ $t$-matrices calculated within this model dynamically generate, in s-wave, the resonance $f_0(980)$, while the system formed by the channels $K\bar K$ and $\pi\eta$ gives rise to the $a_0(980)$ state.

Once the two-body amplitudes for the different subsystems are determined we can study the $KK\bar K$ system. To do that, we consider the set of coupled channels given by $KK\bar K$, $K\pi\pi$ and $K\pi\eta$ and obtain the three-body $T$ matrix for the transitions between the different channels using the formalism developed in 
Refs.~\cite{mko1,mko2,mko3,mko7}, which is based on the Faddeev equations \cite{Fa}.  
The calculation of the three-body $T$ matrix is done for real values of the three-body
energy and peaks  found in the modulus squared  of the three-body $T$-matrix can be associated with dynamically generated resonances.

In this approach, the Faddeev partitions, $T^1$, $T^2$ and $T^3$,  are written as \cite{mko1,mko2,mko7}
\begin{equation}
T^i =t^i\delta^3(\vec{k}^{\,\prime}_i-\vec{k}_i) + \sum_{j\neq i=1}^3T_R^{ij}, 
\label{Ti}
\end{equation}
for $i=1,2,3$ with $\vec{k}_{i}$ ($\vec{k}^\prime_{i}$) being the initial (final) momentum of the particle $i$ and $t^{i}$, $i=1,2,3$, the two-body $t$-matrix that describes the interaction for the $(jk)$ pair of the system, $j \neq k\neq i=1,2,3$,
and which is calculated as explained above. 

\begin{figure}
\includegraphics[scale=0.7]{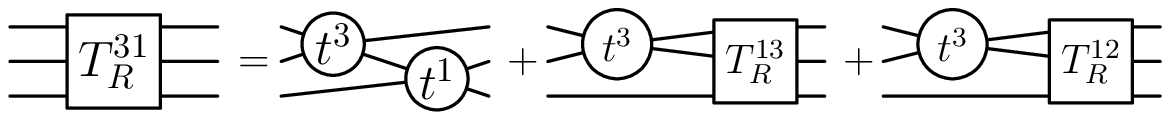}
\caption{Schematic representation of Eq.~(\ref{Trest}) for the partition $T_{R}^{31}$.}
\label{fig:sche}
\end{figure}

The $T^{ij}_{R}$ partitions, which contain all the different contributions to the three-body $T$  matrix in which the last two interactions are given in terms of the two-body $t$-matrices $t^j$ and $t^i$, respectively, satisfy the following set of coupled equations,
\begin{equation}
T^{\,ij}_R = t^ig^{ij}t^j+t^i\Big[G^{\,iji\,}T^{\,ji}_R+G^{\,ijk\,}T^{\,jk}_R\Big], 
  \label{Trest}
\end{equation}
for  $i\ne j, j\ne k = 1,2,3$. 
A schematic representation of Eq.~(\ref{Trest}) is given in Fig.~\ref{fig:sche}.

In Eq.~(\ref{Trest}), $g^{ij}$ correspond to the three-body Green's function of the system and its elements are defined as
\begin{eqnarray}
\lefteqn{
g^{ij} (\vec{k}^\prime_i, \vec{k}_j)=\Bigg(\frac{N_{k}}{2E_k(\vec{k}^\prime_i+\vec{k}_j)}\Bigg)}&& \nonumber \\
&& \times \frac{1}{\sqrt{s}-E_i
(\vec{k}^\prime_i)-E_j(\vec{k}_j)-E_k(\vec{k}^\prime_i+\vec{k}_j)+i\epsilon},
\end{eqnarray}
with $N_{k}=1$ for mesons  and $E_{l}$, $l=1,2,3$, is the energy of the particle $l$.

The $G^{ijk}$ matrix in Eq.~(\ref{Trest}) represents  a loop function of three-particles and it is written as
\begin{equation}
G^{i\,j\,k}  =\int\frac{d^3 k^{\prime\prime}}{(2\pi)^3}\tilde{g}^{ij} \cdot F^{i\,j \,k}
\label{eq:Gfunc}
\end{equation}
with the elements of  $\tilde{g}^{ij}$ being 
\begin{eqnarray}
\lefteqn{
\tilde{g}^{ij} (\vec{k}^{\prime \prime}, s_{lm}) = \frac{N_l}
{2E_l(\vec{k}^{\prime\prime})} \frac{N_m}{2E_m(\vec{k}^{\prime\prime})} } && \nonumber \\
&& \quad \times
\frac{1}{\sqrt{s_{lm}}-E_l(\vec{k}^{\prime\prime})-E_m(\vec{k}^{\prime\prime})
+i\epsilon},
\label{eq:G} 
\end{eqnarray}
for $i \ne l \ne m$, 
and the matrix $F^{i\,j\,k}$, with explicit variable dependence, is given by 
\begin{eqnarray}
\lefteqn{
F^{i\,j\,k} (\vec{k}^{\prime \prime},\vec{k}^\prime_j, \vec{k}_k,  s^{k^{\prime\prime}}_{ru})=  } && \nonumber \\
&& t^{j}(s^{k^{\prime\prime}}_{ru}) g^{jk}(\vec{k}^{\prime\prime}, \vec{k}_k)
\Big[g^{jk}(\vec{k}^\prime_j, \vec{k}_k) \Big]^{-1}
\Big[ t^{j} (s_{ru}) \Big]^{-1},  \label{offac}
\end{eqnarray}
for $ j\ne r\ne u=1,2,3$.
In Eq. (\ref{eq:G}), $\sqrt{s_{lm}}$ is the invariant mass of the $(lm)$ pair and can be calculated in terms of the external variables. The upper index $k^{\prime\prime}$ in the invariant mass $s^{k^{\prime\prime}}_{ru}$ of Eq.~(\ref{offac}) indicates its dependence on the loop variable, as it was shown in Ref. \cite{mko2}.

Equation~(\ref{Trest}) is an algebraic coupled equation since it involves
only the on-shell part of the two-body $t$-matrices. This is due to the finding  of  a  cancellation between the contribution of the off-shell parts of the chiral two-body $t$-matrices to the three-body Faddeev amplitudes and a contact term with same topology  whose origin is in the chiral Lagrangian used to describe the interaction. It was found in the three-body system of two pseudoscalar mesons and one baryon or one vector meson that this cancellation becomes exact in the flavor SU(3) limit~\cite{mko1,mko2,mko4}\footnote{Although in Ref. \cite{mko1} the cancellation was found to be exact in the SU(3) limit and assuming a small momentum transfer for the baryon, this last condition was shown to be unnecessary in Refs. \cite{mko2,mko4} for a two meson and one baryon system and a three meson system with one of the mesons being a vector meson.}, and that in a realistic case off the SU(3) limit the sum of the off-shell part and the three-body contact term was estimated to be smaller than $5\%$ of the total on-shell contribution. Thus, only the on-shell part of the two-body (chiral) $t$-matrices was significant. For the present case of a three pseudoscalar meson system, as we  show in the Appendix, an exact analytic cancellation can be achieved in the chiral limit by considering two more diagrams which involve $s$-channel intermediate states of one meson and five mesons\footnote{Note that  for the cancellation found in Refs.~\cite{mko1,mko2,mko4}, due to the particular structure of the pseudoscalar-baryon and pseudoscalar-vector Weinberg-Tomozawa interactions, these diagrams can only contribute to the on-shell terms and, thus, they were not necessary for the cancellation.}. 
In a realistic calculation off the chiral limit, we find numerically that the breaking of the cancellation accounts for about $7\%$ of the total on-shell contribution. This result is very similar to the one found in Refs. \cite{mko1,mko2,mko4} when the SU(3) limit was not taken. Therefore, for our purpose, we can neglect the contribution coming from the off-shell parts of the two-body chiral $t$-matrices, the corresponding contact term from the chiral Lagrangian and the diagrams with one and five meson intermediate states, and work only with the on-shell part of the two-body $t$-matrices and  three-body intermediate states.

The $T^{ij}_R$ partitions given in Eq.~(\ref{Trest}) are calculated as a function of the total three-body energy, $\sqrt s$, and the 
invariant mass of  the particles 2 and 3, $\sqrt{s_{23}}$. The other invariant masses, $\sqrt{s_{12}}$ and $\sqrt{s_{31}}$ can be obtained in terms of  $\sqrt s$ and $\sqrt{s_{23}}$, as it was shown in Ref. \cite{mko2,mko3}.
To present our results, we have chosen $\sqrt{s}$ and the invariant mass of the (23) two-body subsystems, $\sqrt{s_{23}}$ .
Because we are interested in generating the s-wave $f_0(980)$ and $a_0(980)$ states in the $K\bar K$ subsystem, we project all the matrices present in  Eq.~(\ref{Trest}) on s-wave, which implies that the quantum numbers of the three-body system and, hence, the resulting bound states or resonances are $J^{\pi}=0^{-}$.

The full three-body $T$-matrix is obtained in terms of the two-body $t$-matrices and 
the $T^{ij}_{R}$  partitions as
\begin{equation}
T = T^{1} + T^{2} + T^{3} = \sum_{i}^{3}t^i\delta^3(\vec{k}^{\,\prime}_i-\vec{k}_i) +T_{R}\label{T}
\end{equation}
where
\begin{equation}
T_{R} \equiv \sum_{i=1}^3\sum_{j\neq i=1}^{3}T^{ij}_{R} . \label{ourfullt}
\end{equation}

In this formulation, the symmetry property for the exchange of two identical particles 
is automatically incorporated to 
the full $T$ and  $T_{R}$ matrices as far as that symmetry is implemented
in the corresponding two-body $t$-matrices, which is the case here.

The two-body $t$-matrices present in Eq.~(\ref{T}) can not give rise to any three-body structure, thus,
to identify possible three-body states one can concentrate in studying the properties of the $T_R$ matrix defined in
Eq.~(\ref{ourfullt}).

\section{Results}\label{Re}

Our interest is to examine the possibility of existence of kaonic states in the $KK\bar K$ system. For this purpose, we solve numerically  Eq.~(\ref{Trest}) in a coupled channel approach considering the following set of coupled channels for total charge zero: $K^0K^+K^-$, $K^0K^0\bar K^0$, $K^0\pi^+\pi^-$, $K^0\pi^-\pi^+$, $K^0\pi^0\pi^0$, $K^0\pi^0\eta$, $K^+K^0K^-$, $K^+\pi^0\pi^-$, $K^+\pi^-\pi^0$, $K^+\pi^-\eta$, and we take
isospin-averaged masses for the different mesons, i.e., $m_K$ (for $K^+, K^-, K^0, \bar{K}^0$), $m_\pi$ (for $\pi^+,\pi^-,\pi^0$) and $m_\eta$. 
To identify the isospin associated with peaks in the $|T_R|^2$, we make an appropriate unitary transformation 
from the charged base to an isospin base  which is characterized by the total isospin of the three-body system, $I$, and the isospin of one of the two-body subsystems, $I_{ab}$. In this way we can understand further the properties of the possible resulting states. Naming the particles in the $K M_{2} M_{3}$ system as 1, 2 and 3, we calculate the transition amplitude in the isospin base $|I,I_{23}\rangle$ as follows
\begin{equation}
T^{(I,I_{23})}_R(\sqrt s, \sqrt{s^{I_{23}}_{23}}) \equiv\langle I, I_{23}|T_R(\sqrt s,\sqrt s_{23})| I,I_{23}\rangle.
\end{equation}
Resonance or bound states  are determined as peaks in the modulus squared of the $T^{(I,I_{23})}_{R}$ amplitude, which depends on the energy of the three-body system, $\sqrt s$, and the invariant mass of the (23) subsystem projected on isospin $I_{23}$, $\sqrt{s^{I_{23}}_{23}}$. The mass and the width of the state is read off from  the peak position of $|T_R|^2$. In general, if resonances are formed due to the three-body dynamics, they
appear in all the coupled channels. We present here the results involving the  $T_{R}$ amplitude for the particular case 
$KK \bar K \to KK \bar K$.

\begin{figure}[ht!]
\centering
\includegraphics[scale=0.6]{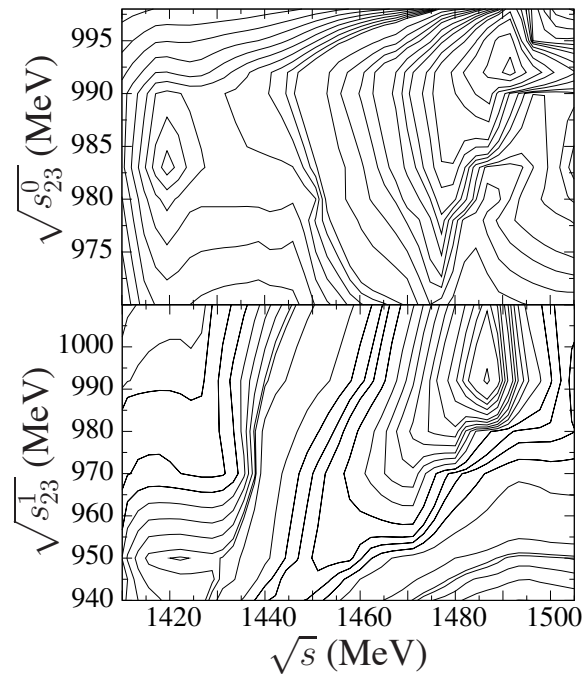}
\caption{Contour plots of the three-body squared amplitudes $\Big|T^{(1/2,0)}_{R}\Big|^2$
and $\Big|T^{(1/2,1)}_{R}\Big|^2$ for the $KK\bar{K} \to KK \bar K$ transition 
with total $I=1/2$ as functions of the total three-body energy, $\sqrt s$, and 
the invariant mass of the $K\bar K$ subsystem with $I_{23}=0$ (upper panel)
or the invariant mass of the $K\bar K$ subsystem with $I_{23}=1$ (lower panel).
} \label{N1}
\end{figure}

In Fig.~\ref{N1}, we show the contour plots associated to the modulus squared three-body 
amplitudes  $T^{(1/2,0)}_{R}$ (upper panel) and $T^{(1/2,1)}_{R}$ (lower panel) for the transition
$KK\bar K \to KK\bar K$ with total isospin $I=1/2$. We consider the cases in which
the two-body (23) subsystem is projected on isospin $I_{23}=0$ (upper panel) or $I_{23}=1$ (lower panel) to
have the possibility of generating the $f_0(980)$ or $a_0(980)$, respectively, in that subsystem.

First of all,  we see in both panels of Fig.~\ref{N1} a peak structure at an energy  around $3 m_K\sim1488$ MeV (with $m_K$ = 496 MeV the kaon mass) which appears when the invariant masses of the respective $K\bar K$ subsystems have a value around $2 m_K$, i.e., their threshold values. If we only considerer $K\pi\pi$ and $K\pi\eta$ as coupled channels, the signal at 1488 MeV is not present. Thus, we conclude that the peak which shows up at 1488 MeV corresponds then to the opening of the three-body $KK\bar K$ threshold.  

Apart from this trivial structure, we find a peak 
at $\sqrt s \sim 1420$ MeV and a width of $\sim$ 50 MeV when
$\sqrt {s_{23}^{0}} \sim 983$ MeV, as shown in the upper panel of Fig.~\ref{N1}.
As it can be seen in the lower panel of Fig.~\ref{N1}, this resonance state also shows up for a value of 
$\sqrt {s_{23}^{1}}$ around $950$ MeV and $\sqrt s \sim 1420$ MeV.
These two peaks correspond to a single state with a mass $\sim $ 1420 MeV which can be interpreted as a quasibound state
of the $KK\bar K$ system with one of the $K\bar K$ pairs forming the $f_0(980)$: The resonance shows up when the invariant mass of the $K\bar K$ pair with isospin zero is close to a value of 983 MeV. This means that the $f_0(980)$ resonance is formed in the subsystem. However, when the $K\bar K$ is projected on $I_{23}=1$, the invariant mass for the $K\bar K$ pair has a value around 950 MeV. This value is not exactly in the region where the $a_0(980)$ gets dynamically generated,  but it is also not very far away\footnote{Although the pole associated with the $a_0(980)$ is localized on the complex plane at a real value of the energy near 1009 MeV, the peak corresponding to this state on the real plane  appears around 980 MeV.},
and the attraction present in the system helps in forming a three-body bound state. 
Therefore, for the state found around a total energy of 1420 MeV, both attractions of $K\bar K$ with isospin 0 (and in which the $f_0(980)$ is generated) and 1 are important to form the three-body quasibound state.

This is a similar situation to the three-body resonance state  $N^{*}(1910)$ found in the $K\bar KN$ system~\cite{jido,mj1}, in which  the state is generated by the attraction in $\bar KN$ and $K\bar K$.

The state obtained in the $KK\bar K$ system and coupled channels can probably correspond to the $K(1460)$ listed by the PDG \cite{pdg} (which is omitted from the summary table) and  observed  in $K\pi\pi$ partial wave analysis, although the width found in this work is much smaller than the two values listed by the PDG, around 250 MeV.  Note, however, that the width obtained within this formalism comes only from s-wave three-body channels and, normally, is smaller than the total width observed for that state to which the two-body decay widths also contribute, even if these two-body channels have a smaller weight in the resonance wave function, as implicitly assumed in our study. For example, the inclusion of  $p$-wave channels, like $\pi K^*(892)$ (a decay channel observed for the $K(1460)$ \cite{pdg}), although it should not be essential to generate the state found \footnote{A dynamically generated resonance, like the one found in this manuscript, can be interpreted as an hadronic-molecular resonance where the hadrons forming it retain their structure. This means that the quarks inside the hadrons do not play an essential role in the formation of this type of states. Such resonances are normally  weakly bound states and, thus, the constituent hadrons have little energy, which implies that the probability of formation of such resonances is higher when the interaction between the hadrons is in s-wave because these hadrons have low momenta.}, could definitively help in increasing the width obtained. However, the poor experimental information available in this energy region for kaonic states suggests  that the values of the widths in Ref.~\cite{pdg} may not be very precise.

To understand further the structure of the resonance found at 1420 MeV, we have studied the effect of the different three-body coupled channels
calculating the three-body $T_{R}$ amplitude taking into account the $KK\bar K$ channel and excluding $K\pi\pi$ and $K\pi\eta$ 
in the three-body space\footnote{To determine the two-body $t$ matrices of the $KK$ and $K\bar K$ systems we continue considering
all the two-body coupled channels, except for the $\eta\eta$ channel as done before.}. 
In this calculation, we get again a peak in the $KK\bar K$ amplitude around $\sqrt s \sim 1420$ MeV for total isospin $I=1/2$ with one of the $K\bar K$ pairs forming the $f_0(980)$ and the other not very far away from the region where the $a_0(980)$ gets generated and with a magnitude in $|T_R|^2$ similar to the result obtained with the full set of  coupled channels. If the $KK\bar K$ channel is excluded and the $K\pi\pi$ and $K\pi\eta$ channels are considered as
coupled channels, the corresponding peak is found around 1450 MeV (a bit higher than the resonance position of the full calculation) for total isospin $1/2$. However its signal is much weaker than that of the full calculation.
This fact suggests that for the state found in the $KK\bar K$ system and coupled channels around 1420 MeV the $f_{0}(980)$ resonance plays an important role in determining the resonance position, because the $\pi\pi$ channel  couples weakly to the $f_{0}(980)$ 
in the two-body dynamics as compared to the $K\bar K$ channel and without the $K\bar K$ channel the $f_{0}(980)$ 
resonance cannot participate strongly in the three-body dynamics. Nevertheless, since there is still some attraction from 
the $\pi\eta$ channel in the (23) subsystem with $I_{23}=1$ to form the $a_{0}(980)$ resonance, one can still find a weak signal for the state,  although the $KK\bar K$ channel 
essentially determines the structure of this resonance.

Finally, it is rewarding to mention that in our calculation of the three-body system with total isospin $I=3/2$, 
which is manifestly exotic and cannot be made up with a $q\bar q $ configuration, we do not find any resonance state in the studied energy region: For total isospin $I=3/2$, the two-body subsystems present in the
$KK\bar K$  system have substantially large isospin 1 components. In the $KK$ subsystem the isospin 1 configuration  is repulsive in nature, while for $K \bar K$,  although the isospin 1 configuration is attractive,  the interaction is not as strong as when the $\pi\eta$ channel is also present to  generate together the $a_{0}(980)$ resonance. Thus, for total isospin $3/2$, in the $KK\bar K$ system, the $K \bar K$ interaction in isospin 1  probably is not strong enough to overcome the repulsion originated from the $KK$ interaction in isospin 1 and form a resonance or a bound state. For this reason, states with total isospin $I=3/2$ are probably hard to be generated dynamically in the three-body system. 

\section{Potential model}\label{sec:pot}
As we have seen in Sec.~\ref{Re}, the state found around an energy of $1420$ MeV in the $KK\bar K$, $K\pi\pi$ and $K\pi\eta$ system couples
more strongly to $KK\bar K$ than to $K\pi\pi$ or $K\pi\eta$. This fact, together with the smallness of the kinetic energy of the kaons in a bound system in comparison  with the kaon mass, makes possible the study of the single channel $KK\bar K$ using a nonrelativistic potential model, like the one developed in Refs.~\cite{jido,enyo}, and compare the results with the ones obtained by solving the Faddeev equations for the 
$KK\bar K$ system and coupled channels.

Following Refs.~\cite{jido,enyo}, we consider $KK\bar K$ as a single channel and 
determine its wave function by solving the Schr\"odinger equation for the Hamiltonian of 
the system, which is given by
\begin{equation}
H=T+V_{KK}(r_1)+V_{K\bar K}(r_2)+V_{K\bar{K}}(r_3), \label{eq:hamiltonian}
\end{equation}
with $T$ the kinetic energy of the system and $V_{KK}$, $V_{K\bar K}$ effective potentials which describe the $KK$ and
$K\bar K$ interactions, respectively. These potentials are written in terms of $\ell$-independent  local potentials as functions of 
the $K$-$K $ and $K$-$\bar K$ distances,  $r_{1}$, $r_{2}$ and $r_{3}$ and we take a Gaussian form for them:
\begin{equation}
V^{I}_{k}(r)=U^{I}_{k}
\exp\left[ -(r/b)^2\right]P_{k}(I),\label{pot2}
\end{equation}
where $k$ denotes $K\bar K$ or $KK$,  $I$ is the isospin of the two kaon system,
and $P_k$ represents the isospin projector. The parameters involved in Eq.~(\ref{pot2})
are the interaction range, $b$, and the strength of the potential, $U_{k}^{I}$.
We consider as values for the parameters those used in Refs.~\cite{jido,enyo} to study 
the $K\bar K N$ and $\bar K \bar K N$ systems, which are:
$U^{I=0,1}_{K\bar K}=-1155-283i$ MeV, $U^{I=1}_{KK} = 313$ MeV 
with $b=0.47$ fm.
The $\bar KK$ interaction strengths were determined  to have a quasibound 
state with mass 980 MeV and width 60 MeV in isospin 0 and isospin 1, which correspond to
the $f_{0}(980)$ and $a_{0}(980)$ resonances, respectively. This means that the attractive $\bar KK$ interactions have the same strengths for both $I_{\bar KK}=0$ and $I_{\bar KK}=1$.
The strength of the repulsive $KK$ interaction in $I_{KK}=1$ was fixed to 
reproduce the scattering length $a_{K^{+}K^{+}}=-0.14$, which has been obtained 
from a lattice QCD calculation~\cite{Beane:2007uh}. Since the $s$-wave interaction 
for $KK$ in isospin 0 is forbidden due to Bose statistics, we consider $U^{I=0}_{KK}=0$. 
In Refs.~\cite{jido,enyo}, another parameter set is given for $b$ and $U^{I}_{k}$.
We have also tried that set of parameters  and obtained very similar results to the ones shown below. 

In solving the Schr\"odinger equation for the $KK\bar{K}$ channel, we first consider 
only the real part of the potentials and determine the corresponding wave functions using 
a variational approach as in Refs.~\cite{jido,enyo}. Then we calculate the bound state energies 
$E$ as the expectation values  of the Hamiltonian defined in Eq.~(\ref{eq:hamiltonian})  
with  respect to the obtained wave functions. 
The binding energy is determined from the real part of the calculated energy,
while the widths of the bound states are evaluated from the imaginary part of 
the complex energies as $\Gamma=-2\, {\rm Im}E$.

\begin{table}[tb]
\begin{tabular}{ccc}
\hline\hline
Model & Faddeev calculation & Potential Model \\
\hline
Mass [MeV] & $\sim$ 1420 & 1467  \\
Width [MeV] & $\sim$ 50 & 110  \\ 
\hline
root mean squared radius  [fm] &  - & 1.6  \\
$K$-$K$ distance [fm] & - &  2.8  \\
$(KK)$-$\bar K$ distance [fm] &  - & 1.7  \\
\hline
$K_{2}$-$\bar K_3$ distance [fm]$^{\dagger}$ & - & 1.6  \\
$K_{1}$-$(K_2\bar K_3)$ distance [fm]$^{\dagger}$ & - & 2.6  \\
\hline\hline
\end{tabular}
\caption{Comparison of the results of the Faddeev calculation and the potential model. The spatial structure of the $KK\bar K$ quasibound state obtained with the potential model is also shown in the table. $^{\dagger}$The values of the $K_{2}$-$\bar K_3$ and $K_{1}$-$(K_2\bar K_3)$ distances 
are obtained before making the symmetrization of $K_{1}K_{2}$.
\label{table} }
\end{table}

We get as a result a quasibound state of the $KK\bar K$ system
with 21 MeV binding energy and 110 MeV width.
This state appears for an energy similar 
to the one of the resonance obtained in the Faddeev calculation of Sec.~\ref{Re}. A comparison of the results found
with both methods is given in Table~\ref{table}. It is interesting to notice that although the two methods 
are very different, the energy position of the quasibound $KK\bar K$ state does not differ very much.
Note, however, that in the potential model used we consider only the single $KK\bar K$ channel and
do not take into account the possible modification of the two-body interaction in the 
presence of the third particle. In such simple calculation, for weakly bound systems, 
the resulting binding energy and width correspond to the sum of the binding energies and widths 
of the two-body subsystems, $f_{0}(980)$ and $a_{0}(980)$ in the present case, as discussed in Ref.~\cite{jido}.

In the potential model, the three-body wave function is also obtained. 
With the wave function we can investigate the spacial structure of the three-body 
quasibound state. 
We obtain the root mean squared radius of the $KK\bar K$ quasibound state 
to be 1.6 fm. This value is  similar  to the one found for the $K\bar KN$ system~\cite{jido}, which was 1.7 fm.
The average $K$-$K$ distance and the distance 
between the $KK$ cluster and $\bar K$ are calculated and found to be 2.8 fm and 
1.7 fm, respectively. 
The distance of the repulsive $KK$ is also very similar to the corresponding 
result for the $KN$ distance in the $K\bar KN$ system. 

\begin{figure}[tb]
\includegraphics[scale=0.4]{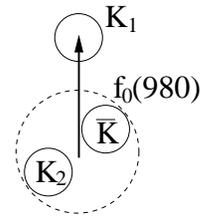}
\caption{Schematic picture of the $KK\bar K$ quasibound state before making
the symmetrization of $K_{1}$ and $K_{2}$. \label{fig:struc}}
\end{figure}

In the calculation, we have taken three Jacobian coordinates of the 
$K_{1}K_{2}\bar K_{3}$ system, $K_{1}$-$(K_{2}\bar K_{3})$, $K_{2}$-$(K_{1}\bar K_{3})$
and $\bar K_{3}$-$(K_{1}K_{2})$.
Even without the $\bar K_{3}$-$(K_{1}K_{2})$ rearrangement channel,
we obtain  almost the same three-body binding energy. This means that 
the system can be described essentially by the 
$K_{1}$-$(K_{2}\bar K_{3})+K_{2}$-$(K_{1}\bar K_{3})$
configuration. Since the $s$-wave configuration 
of $K_{1}K_{2}$ with $I_{12}=0$ is not allowed because of Bose statistics, 
the $K_{1}K_{2}$ subsystem has almost $I_{12}=1$. In case that the $K_{1}K_{2}$ subsystem
has purely $I_{12}=1$, according with the isospin algebra, the $K \bar K$ subsystem should be composed by isospin 0 and 1 
in a ratio of 3:1 for total isospin $I=1/2$ of the $KK\bar K$ system. 
Calculating the distances of $K_{2}$-$\bar K_{3}$ and $K_{1}$-$(K_{2}\bar K_{3})$
without the symmetrization of $K_{1}$ and $K_{2}$, we obtain 
1.6 fm and 2.6 fm, respectively. A schematic picture is given in Fig.~\ref{fig:struc}. 
These facts indicate that 
the $KK\bar K$ quasibound system can be interpreted as a system 
of one $K$ and a strongly correlated $K \bar K$ pair forming dominantly 
the $f_{0}(980)$. This feature is consistent with what we have found 
for the $KK\bar K$ resonant state in the Faddeev calculation. 
It is also worth noting
that the attraction of $K\bar K$ in isospin 1 is essential
to make the interaction between $K$ and $f_0(980)$ attractive
enough to form a quasibound state.
Actually when we do not include attraction in $K \bar K$ with isospin 1,
we find no bound state for the $KK\bar K$ system.

Here we emphasize an important role of the repulsive 
$KK$ interaction for the hadronic molecular states. One of the crucial 
features of the hadronic molecular state is that the system can be 
described by its hadronic constituents. Therefore, if a quasibound state 
is formed in  the hadronic system, it should be such a loosely bound state
that the constituents hadrons are separated and keeping their identities inside 
the bound state. For deeply bound states, such as ones with hundreds of
MeV binding energy, the constituent hadrons get close to each other 
and may be overlapped. In such a case, the hadronic molecular picture 
may be broken down for these states and, if such states exist in nature one has to interpret them by other mechanisms than the hadronic molecular picture.  
In the present $KK\bar K$ system, $KK$ with isospin 1 has a 
repulsion, and this keeps the $KK\bar K$ system loosely bound with a moderate binding energy.
When we take an artificial attraction for the $KK$ interaction as strong as the $K\bar K$ interaction, 
we obtain a 
very deeply bound state with hundreds of MeV binding energy. Thus,
for  a hadronic molecular state appearing near the threshold of the system, it is necessary 
that one of the pairs has a repulsive interaction or, at least, sufficiently weak 
attraction.

Because in this calculation we have considered the same potential for the cases $I_{K\bar K}=0$ and $I_{K\bar K}=1$, for total isospin  $I=3/2$ we obtain a quasibound state at the same energy as the one found for total isospin $I=1/2$. However, the origin of this state is in the
use of an isospin independent $K\bar K$ interaction in the potential model: Although the $f_0(980)$ and $a_0(980)$ resonances can be considered as bound states
of the $K\bar K$ system for isospin 0 and 1, respectively, at almost the same energy, one should also consider some non-resonant contributions, as done in the Faddeev calculation, which certainly makes the $K\bar K$ interaction in isospin 0 different from the one in isospin 1, resulting, as shown in Sec.~\ref{Re}, in no quasibound state in the $KK\bar K$ system for total isospin 3/2.

\section{Conclusions}\label{Co}
We have studied the $KK\bar K$ system and coupled channels solving the Faddeev equations
within an unitary chiral approach to describe the interaction between the different subsystems.
A resonance state around 1420 MeV, thus below the $KK\bar K$ threshold, which couples dominantly to the $KK\bar K$ channel, is found when one of
the $K\bar K$ pairs is in isospin zero generating the resonance $f_0(980)$ and the other one
is in isospin 1, not very far from the region where the $a_0(980)$ is generated.
A non-relativistic potential model has been
also employed to study the $KK\bar K$ channel and similar results to those of the Faddeev approach has been obtained.
A state of these characteristics could be probably observed in processes involving a final state of three
particles like $KK\bar K$, as an enhancement of the cross section close to threshold due to the presence of the quasibound  $KK\bar K$ state below it, by taking coincidence of the $f_0(980)$ 
out of the invariant masses of the $K\bar K$ pairs. And also in processes involving final states
like $K\pi\pi$ or $K\pi\eta$, although the coupling to these channels is much smaller and it can be more difficult to find the state.

The resonance state obtained in the $KK\bar K$ system can be an analog state
of the quasibound states found in $K\bar KN$~\cite{jido,mko3,mj1}, 
in $\bar KNN$~\cite{dote} and in $\bar K \bar KN$~\cite{enyo}. 
These states are found to be loosely bound in the systems of kaons and nucleons. 
The significant similarity among these systems stems from the fact that 
the two-body $K \bar K$ and $\bar KN$ systems  with isospin 0 
have sufficient attraction to form loosely bound states with a dozen MeV binding
energy. This is because, although kaons and nucleons, having different masses, are kinematically different,
the fundamental interactions among $K\bar K$ and 
$\bar KN$ in $s$-wave at low energy are determined by current algebra 
as a consequence of the spontaneous breaking of the chiral symmetry in QCD,
and the strengths of these interactions are identical in the sense 
of the SU(3) flavor symmetry being enough to make a bound state~\cite{Hyodo:2006yk}. 
Even though the attractions in the two-body subsystems are important 
to form three-body quasibound states in kaonic nuclear systems,
some repulsive interaction or sufficiently weak attraction 
in one of the two-body subsystems is necessary to form a hadronic 
molecular state.

\begin{acknowledgments}
The work of A.~M.~T.~is supported by  
the Grant-in-Aid for the Global COE Program ``The Next Generation of Physics, 
Spun from Universality and Emergence" from the Ministry of Education, Culture, 
Sports, Science and Technology (MEXT) of Japan.
This work is supported in part by
the Grant for Scientific Research (No.~22105507 and No.~22540275) from 
MEXT of Japan.
A part of this work was done in the Yukawa International Project for 
Quark-Hadron Sciences (YIPQS). One of the authors, A. M. T, thanks K. P. Khemchandani
for useful discussions.
\end{acknowledgments}

\appendix*
\section{Cancellation of the off-shell contribution}
\begin{figure*}
\centering
\includegraphics[width=\textwidth]{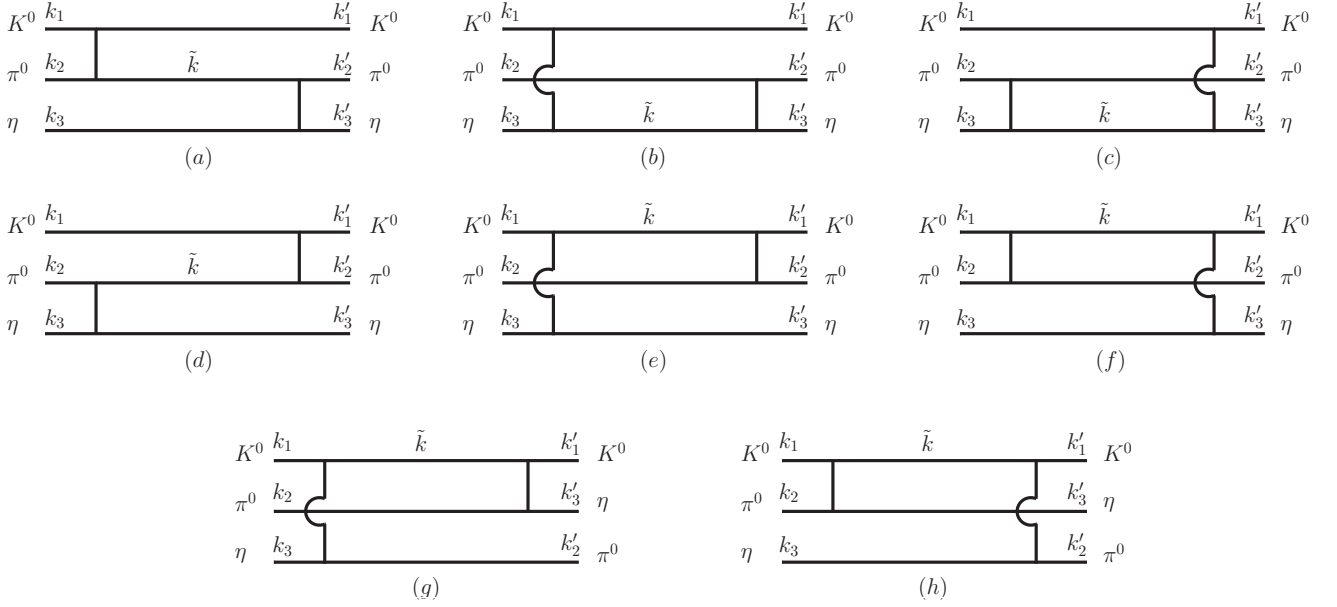}
\caption{Faddeev diagrams contributing to the process $K^0\pi^0\eta\to K^0\pi^0\eta$ with three-meson intermediate states. The vertical lines represent the tree level amplitudes given in Eqs.~(\ref{V1})-(\ref{V4}).}\label{Fdiag}
\end{figure*}

In Ref.~\cite{mko2}  a cancellation between the contribution of the off-shell part of the two-body
chiral $t$-matrices to the Faddeev diagrams and a contact term whose origin stands on the chiral Lagrangian used to describe the interaction
between the particles
has been shown to be exact in the 
flavour SU(3) limit for a system made of two pseudoscalar mesons and a baryon. In Ref.~\cite{mko4} the same type of cancellation
was obtained for a system of three mesons, one of them being a vector meson. In this appendix we are going to show that 
in a three pseudoscalar system a cancellation similar to that  mentioned above can be found and  this cancellation turns out to be exact in the chiral limit in which the mesons are assumed to be massless. 
To do that, we consider, as an example, the process $K^0\pi^0\eta\to K^0\pi^0\eta$. In Fig. \ref{Fdiag}
we show the Faddeev diagrams contributing to this process taking into account all the possible three-meson intermediate states present in the process under
consideration. For the proof of the cancellation, it is sufficient to consider the lowest order diagrams, since the rescattering effects can be factored out. 

To calculate the contribution of the diagrams shown in Fig. \ref{Fdiag}, we need to know the two-body amplitudes $K^0\pi^0\to K^0\pi^0$, $K^0\pi^0\to K^0\eta$, $K^0\eta\to K^0\eta^0$ and $\pi^0\eta\to\pi^0\eta$. At lowest order, the chiral Lagrangian which describes the interaction between any number of pseudoscalar mesons (P) is given by
\begin{equation}
\mathcal{L}=\frac{f^2}{4}\langle \partial_\mu U^\dagger\partial_\mu U +M(U+U^\dagger)\rangle,\label{L}
\end{equation}
where $f$ is the pion decay constant and $\langle \,\rangle$ stands for the trace of the matrices built out of $U(\Phi)$ and $M$,
with
\begin{equation}
U(\Phi)=e^{i\sqrt{2}\Phi/f},
\end{equation}
and $\Phi$ a matrix containing the different Goldstone boson fields,
\begin{align}
\Phi=\left(\begin{array}{ccc}\frac{1}{\sqrt{2}}\pi^0+\frac{1}{\sqrt{6}}\eta & \pi^+ & {K}^+ \\ \pi^- & -\frac{1}{\sqrt{2}}\pi^0+\frac{1}{\sqrt{6}}\eta & {K}^0 \\{K}^- & \bar{K}^{0} & -\frac{2}{\sqrt{6}}\eta\end{array}\right),\label{phi}
\end{align}
and $M$ the mass matrix,
\begin{align}
M=\left(\begin{array}{ccc}m^2_\pi &0&0\\0& m^2_\pi& 0\\0&0&2m^2_K-m^2_\pi\end{array}\right).\label{mass}
\end{align}
The $\eta$ meson mass is given by the Gell-Mann--Okubo mass relation
\begin{equation}
m_{\eta}^{2} = \frac{1}{3}(4m_{K}^{2}-m_{\pi}^{2}).\label{Okubo}
\end{equation}

If we expand $U$ in series up to terms containing four pseudoscalar fields $\Phi$, Eq. (\ref{L}) becomes
\begin{equation}
\mathcal{L}_{4P} =\frac{1}{12f^2}\langle (\partial_\mu \Phi \Phi-\Phi\partial_\mu\Phi)^2+M\phi^4\rangle.
\end{equation}
Using this Lagrangian, we get
\begin{align}
V_{K^0 \pi^0\to K^0\pi^0}&=\frac{1}{12 f^2}[s-2t+u-2m^2_K-2m^2_\pi]\label{V1}\\
V_{K^0 \pi^0\to K^0\eta}&=-\frac{1}{12\sqrt{3}f^2}[3(s-2t+u)\nonumber\\
&\quad+2m^2_K-2m^2_\pi]\label{V2}\\
V_{K^0 \eta\to K^0\eta}&=\frac{1}{12f^2}[3(s-2t+u)\nonumber\\
&\quad - 6m^2_K+2m^2_\pi]\label{V3}\\
V_{\pi^0\eta\to\pi^0\eta}&=-\frac{m^2_\pi}{3f^2},\label{V4}
\end{align}
with the Mandelstam variables $s=(k_1+k_2)^2$, $t=(k_1-k_3)^2$ and $u=(k_1-k_4)^2$ for a process $P_1 P_2\to P_3 P_4$ with
$k_i$ the four momenta of the external particles.
Considering the identity
$s+t+u=\sum k^2_i$, we can write these amplitudes as
\begin{align}
V_{K^0 \pi^0\to K^0\pi^0}&=\frac{1}{12 f^2}[-3t+\sum_i (k^2_i-m^2_i)]\label{Kpi}\\
V_{K^0 \pi^0\to K^0\eta}&=-\frac{1}{12\sqrt{3}f^2}[-9t+8m^2_K+m^2_\pi\nonumber\\
&\quad+3m^2_\eta+3\sum_i (k^2_i-m^2_i)]\label{tran}\\
V_{K^0 \eta\to K^0\eta}&=\frac{1}{12f^2}[-9t+6m^2_\eta\nonumber\\
&\quad+2m^2_\pi+3\sum_i (k^2_i-m^2_i)],\label{Keta}
\end{align}
with $m_i$ the mass of the external particles and $t$ given in terms of the external momenta involved in the two-body amplitudes  of the diagrams shown in Fig.~\ref{Fdiag}.

Using Eqs.~(\ref{Kpi}), (\ref{Keta}), we calculate, for example, 
the contribution of the diagram Fig.~\ref{Fdiag}e as
\begin{align}
T^{(e)}&=\frac{1}{144f^4}[-9{\Delta k_3}^2+6m^2_\eta+2m^2_\pi+3(\tilde k^2-m^2_K)]\nonumber\\
&\quad\times\frac{1}{\tilde k^2-m^2_K}[-3{\Delta k_2}^2+(\tilde k^2-m^2_K)]\nonumber\\
&\equiv T^{(e)}_{on}+T^{(e)}_{off},
\end{align}
where $\Delta k_i\equiv k_i-k^\prime_i$ ($i=1,2,3$) and with $T^{(e)}_{on}$ ($T^{(e)}_{off}$) the contribution which comes from the on-shell (off-shell) part of the amplitudes:
\begin{align}
T^{(e)}_{on}&=-\frac{1}{48f^4}[-9{\Delta k_3}^2+6m^2_\eta+2m^2_\pi]\nonumber\\
&\quad\times\frac{{\Delta k_2}^2}{{\Delta k_2}^2-2k^\prime_1\Delta k_2}\nonumber\\
T^{(e)}_{off}&=\frac{1}{144f^4}[-9{\Delta k_3}^2-6{\Delta k_2}^2-6k^\prime_1\Delta k_2\nonumber\\
&\quad+6m^2_\eta+2m^2_\pi].\label{Te}
\end{align}
For the rest of diagrams in Fig.~\ref{Fdiag}, analogously to Eq.~(\ref{Te}), 
we obtain their off-shell parts:
\begin{align}
T^{(a)}_{off}&=-\frac{1}{36f^4} m^2_\pi\label{Ta}\\
T^{(b)}_{off}&=-\frac{1}{12f^4} m^2_\pi\label{Tb}\\
T^{(c)}_{off}&=-\frac{1}{12f^4} m^2_\pi\label{Tc}\\
T^{(d)}_{off}&=-\frac{1}{36f^4} m^2_\pi\label{Td}\\
T^{(f)}_{off}&=\frac{1}{144f^4}[-9{\Delta k_3}^2-6{\Delta k_2}^2+6k_1\Delta k_2\nonumber\\
&\quad+6m^2_\eta+2m^2_\pi]\label{Tf}
\end{align}
\newpage
\begin{align}
T^{(g)}_{off}&=\frac{1}{144f^4}\Bigg[-9(k_3-k^\prime_2)^2-9(k_2-k^\prime_3)^2\nonumber\\
&\quad+\frac{3}{2}\Bigg\{(k_1+k_3-k^\prime_2)^2+(k^\prime_1+k^\prime_3-k_2)^2\Bigg\}\label{Tg}\\
&\quad+13m^2_K+2m^2_\pi+6m^2_\eta\Bigg]\nonumber\\
T^{(h)}_{off}&=\frac{1}{144f^4}\Bigg[-9(k_2-k^\prime_3)^2-9(k_3-k^\prime_2)^2\nonumber\\
&\quad+\frac{3}{2}\Bigg\{(k_1+k_2-k^\prime_3)^2+(k^\prime_1+k^\prime_2-k_3)^2\Bigg\}\label{Th}\\
&\quad+13m^2_K+2m^2_\pi+6m^2_\eta\Bigg],\nonumber
\end{align}
where in Eq.~(\ref{Tg}) and Eq.~(\ref{Th}) we have used that $k_1+k_3-k^\prime_2=k^\prime_1+k^\prime_3-k_2$ and $k_1+k_2-k^\prime_3=k^\prime_1+k^\prime_2-k_3$, respectively.
\begin{figure*}[t!]
\centering
\includegraphics[width=\textwidth]{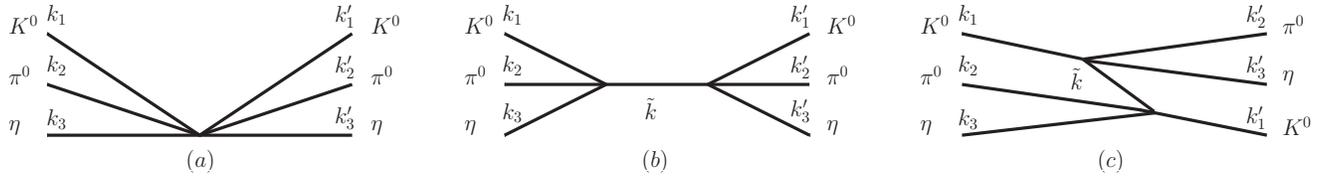}
\caption{Contact term whose origin stands on the Lagrangian of Eq.~(\ref{L}) (a) and
terms with one-meson and five-meson intermediate states (b and c)  contributing to the process $K^0\pi^0\eta\to K^0\pi^0\eta$.}\label{contact}
\end{figure*}

In accordance with the findings of Refs.~\cite{mko1,mko2,mko4}, the contribution of the off-shell part for the different diagrams of Fig. \ref{Fdiag}, together
with the corresponding three pseudoscalar contact terms of the chiral Lagrangian (see Fig.~\ref{contact}b), is expected to vanish under some limit. In  case of Refs.~\cite{mko1,mko2,mko4},
the cancellation found was exact in the SU(3) limit. For this case, a three pseudoscalar system, 
we show that the cancellation is exact under the chiral limit. It is interesting to notice that for a three-pseudoscalar system, apart from the mentioned contact term, we can have two more diagrams which involve one-meson and five-meson intermediate states with the same initial and final states as those shown in Fig.~\ref{Fdiag} (see Fig.~\ref{contact}b and Fig.~\ref{contact}c).   In the energy range of interest for the three-body quasibound state, 1300 -1500 MeV, the on-shell contributions of the diagrams Fig.~\ref{contact}b and Fig.~\ref{contact}c are negligibly small, since they have large energy denominators in the intermediate states.

Let us now evaluate the contribution from the contact term  of Fig.~\ref{contact}a. To do that we need to expand the Lagrangian of Eq.~(\ref{L}) up to terms involving 6 pseudoscalar meson fields, obtaining 
\begin{widetext}
\begin{align}
\mathcal{L}_{6P}&=\frac{1}{360f^4}\langle -9\partial_\mu\Phi\Phi\partial^\mu\Phi\Phi^3+11\partial_\mu\Phi\Phi^2\partial^\mu\Phi\Phi^2-4\partial_\mu\Phi\Phi^3\partial^\mu\Phi\Phi+2\partial_\mu\Phi\Phi^4\partial^\mu\Phi-4\Phi\partial_\mu\Phi\Phi^3\partial^\mu\Phi\nonumber\\
&\quad+11\Phi^2\partial_\mu\Phi\Phi^2\partial^\mu\Phi-9\Phi^3\partial_\mu\Phi\Phi\partial^\mu\Phi+6\partial_\mu\Phi\partial^\mu\Phi\Phi^4+6\Phi^4\partial_\mu\Phi\partial^\mu\Phi-15\Phi\partial_\mu\Phi\partial^\mu\Phi\Phi^3+5\Phi\partial_\mu\Phi\Phi\partial^\mu\Phi\Phi^2\nonumber\\
&\quad-10\Phi\partial_\mu\Phi\Phi^2\partial^\mu\Phi\Phi+5\Phi^2\partial_\mu\Phi\Phi\partial^\mu\Phi\Phi-15\Phi^3\partial_\mu\Phi\partial^\mu\Phi\Phi+20\Phi^2\partial_\mu\Phi\partial^\mu\Phi\Phi^2-2M\Phi^6\rangle.\label{L6P}
\end{align}

\end{widetext}
Taking into account Eq.~(\ref{phi}) particularized for the process $K^0\pi^0\eta\to K^0\pi^0\eta$ and Eq.~(\ref{mass}), ~Eq.~(\ref{L6P}) adopts the form
\clearpage
\begin{widetext}
\begin{align}
\mathcal{L}_a&=-\frac{1}{180f^4}(m^2_K+m^2_\pi)K^0\bar K^0\pi^0\pi^0\eta\eta+\frac{1}{120f^4}\Bigg[K^0\bar K^0\partial_\mu\pi^0\partial^\mu\pi^0\eta\eta
+6\partial_\mu K^0\partial^\mu \bar K^0\pi^0\pi^0\eta\eta\nonumber\\
&\quad-3K^0 \partial_\mu \bar K^0\partial^\mu\pi^0\pi^0\eta\eta-3\partial_\mu K^0\bar K^0\partial^\mu\pi^0\pi^0\eta\eta-3K^0\partial_\mu\bar K^0\pi^0\pi^0\partial^\mu\eta\eta
-3\partial_\mu K^0\bar K^0\pi^0\pi^0\partial^\mu\eta\eta\nonumber\\
&\quad+4K^0\bar K^0\pi^0\partial_\mu\pi^0\partial^\mu\eta\eta+K^0\bar K^0\pi^0\pi^0\partial_\mu\eta\partial^\mu\eta\Bigg].\label{LKpieta}
\end{align}
\end{widetext}

Using Eq.~(\ref{LKpieta}) and taking into account that 
\begin{equation}
\Delta k_1+\Delta k_2+\Delta k_3=0\label{conser}, 
\end{equation}
we get for the diagram of Fig.~\ref{contact}a the contribution
\begin{equation}
t^{(a)}_3=\frac{1}{6f^4}{\Delta k_1}^2-\frac{1}{90f^4}(16m^2_K+3m^2_\eta+m^2_\pi).\label{t3a}
\end{equation}

The contribution of the diagrams in Figs.~\ref{contact}b and Figs.~\ref{contact}c can be calculated using the amplitude of Eq.~(\ref{tran}). In particular, for the diagram
in Fig.~\ref{contact}b we have
\begin{align}
t^{(b)}_3&=\frac{1}{144\cdot 3f^4}\Bigg[-9(k_2+k_3)^2+8m^2_K+m^2_\pi+3m^2_\eta\nonumber\\
&\quad+3(\tilde k^2-m^2_K)]\frac{1}{\tilde k^2-m^2_K}\Bigg[-9(k^\prime_2+k^\prime_3)^2\nonumber\\
&\quad+8m^2_K+m^2_\pi+3m^2_\eta+3(\tilde k^2-m^2_K)\Bigg]\nonumber\\
&\equiv t^{(b)}_{3\,on}+t^{(b)}_{3\,off}.
\end{align}
We are interested only in $t^{(b)}_{3\,off}$, which is given by
\begin{align}
t^{(b)}_{3\,off}&=\frac{1}{144f^4}\Bigg[-9(k_2+k_3)^2-9(k^\prime_2+k^\prime_3)^2\nonumber\\
&\quad+\frac{3}{2}\Bigg\{(k_1+k_2+k_3)^2+(k^\prime_1+k^\prime_2+k^\prime_3)^2\Bigg\}\nonumber\\
&\quad+13m^2_K+2m^2_\pi+6m^2_\eta\Bigg],\label{t3boff}
\end{align}
where we have used the fact that $k_1+k_2+k_3=k^\prime_1+k^\prime_2+k^\prime_3$.
Similarly, for the diagram in Fig.~\ref{contact}c we obtain
the off-shell part as
\begin{align}
t^{(c)}_{3\,off}&=\frac{1}{144f^4}\Bigg[-9(k^\prime_2+k^\prime_3)^2-9(k_2+k_3)^2\nonumber\\
&\quad+\frac{3}{2}\Bigg\{(k_1-k^\prime_2-k^\prime_3)^2+(k^\prime_1-k_2-k_3)^2\Bigg\}\nonumber\\
&\quad+13m^2_K+2m^2_\pi+6m^2_\eta\Bigg],\label{t3coff}
\end{align}
where we make use that $k_1-k^\prime_2-k^\prime_3=k^\prime_1-k_2-k_3$.

Summing Eqs.~(\ref{Te}-\ref{Tf}) and using Eq.~(\ref{conser}) we get
\begin{align}
\sum_{i=a}^{f} T^{(i)}_{off}&=-\frac{1}{8f^4}{\Delta k_1}^2+\frac{5}{24f^4}\Delta k_2\Delta k_3\nonumber\\
&\quad+\frac{1}{36f^4}(3m^2_\eta-7m^2_\pi).\label{Tpartialoff}
\end{align}
If we add Eq.~(\ref{t3a}) and Eq.~(\ref{Tpartialoff}) we obtain
\begin{align}
\sum_{i=a}^{f} T^{(i)}_{off}+t^{(a)}_3&=\frac{1}{24f^4}({\Delta k_1}^2+5\Delta k_2\Delta k_3)\nonumber\\
&\,\,-\frac{1}{180f^4}(32m^2_K-9m^2_\eta+37m^2_\pi).\label{one}
\end{align}
Adding now Eqs.~(\ref{Tg}), (\ref{Th}), (\ref{t3boff}) and (\ref{t3coff}) we find
\begin{widetext}
\begin{align}
\sum_{i=g}^{h} T^{(i)}_{off}+\sum_{i=b}^{c} t^{(i)}_{3\,off}&=\frac{1}{24f^4}\Big[-10m^2_\eta-10m^2_\pi+2m^2_K+5k_3k^\prime_2+5k_2k^\prime_3-5k_2k_3-5k^\prime_2k^\prime_3\nonumber\\
&\quad+\Delta k_1(\Delta k_2+\Delta k_3)\Bigg]+\frac{1}{36f^4}(13m^2_K+2m^2_\pi+6m^2_\eta)\nonumber\\
&=\frac{1}{24f^4}\Big[-10m^2_\eta-10m^2_\pi+2m^2_K-5\Delta k_2\Delta k_3-{\Delta k_1}^2\Bigg]+\frac{1}{36f^4}(13m^2_K+2m^2_\pi+6m^2_\eta),
\end{align}
\end{widetext}
which can be reduced to
\begin{align}
\sum_{i=g}^{h} T^{(i)}_{off}&+\sum_{i=b}^{c} t^{(i)}_{3\,off}=-\frac{1}{24f^4}\Big[{\Delta k_1}^2+5\Delta k_2\Delta k_3\Bigg]\nonumber\\
&\quad\quad+\frac{1}{36f^4}(16m^2_K-13m^2_\pi-9m^2_\eta).\label{two}
\end{align}
 
Therefore, summing all the contributions, which is obtained by adding Eq.~(\ref{one}) and Eq.~(\ref{two}), we find that
the term depending on $\Delta k_i$ gets cancel and there is a mass term remaining, which, using Eq.~(\ref{Okubo}), reads as
\begin{equation}
\sum_{i=a}^{h} T^{(i)}_{off}+\sum_{i=a}^{c} t^{(i)}_{3\,off}=-\frac{m^2_\pi}{2 f^4},
\end{equation}
which vanishes in the chiral limit. In this way we obtain 
an exact cancellation in the chiral limit.

\end{document}